\documentclass[a4paper]{spie}  

\usepackage{amsmath,amsfonts,amssymb}
\usepackage{graphicx}
\usepackage[colorlinks=true, allcolors=blue]{hyperref}

\title{JASMINE image simulator for\linebreak high-precision astrometry and photometry}

\author[a]{Takafumi~Kamizuka}
\author[b]{Hajime~Kawahara}
\author[c]{Ryou~Ohsawa}
\author[b]{Hirokazu~Kataza}
\author[c,d]{Daisuke~Kawata}
\author[e]{Yoshiyuki~Yamada}
\author[f]{Teruyuki~Hirano}
\author[c]{Kohei~Miyakawa}
\author[g]{Masataka~Aizawa}
\author[f]{Masashi~Omiya}
\author[c]{Taihei~Yano}
\author[b,c]{Ryouhei~Kano}
\author[c]{Takehiko~Wada}
\author[h]{Wolfgang~L\"{o}ffler}
\author[h]{Michael~Biermann}
\author[c]{Pau~Ramos}
\author[b]{Naoki~Isobe}
\author[b]{Fumihiko~Usui}
\author[c,i]{Kohei~Hattori}
\author[j]{Satoshi~Yoshioka}
\author[k,l]{Takayuki~Tatekawa}
\author[m]{Hideyuki~Izumiura}
\author[c]{Akihiko~Fukui}
\author[c]{Makoto~Miyoshi}
\author[c]{Daisuke~Tatsumi}
\author[c,n]{Naoteru~Gouda}
\affil[a]{
Institute of Astronomy, Graduate School of Science, the University of Tokyo,
2-21-1 Osawa, Mitaka, Tokyo 181-0015, Japan
}
\affil[b]{
Institute of Space and Astronautical Science, Japan Aerospace Exploration Agency,
3-1-1 Yoshinodai, Chuo-ku, Sagamihara, Kanagawa 252-5210, Japan
}
\affil[c]{
National Astronomical Observatory of Japan,
2-21-1 Osawa, Mitaka, Tokyo 181-8588, Japan
}
\affil[d]{
Mullard Space Science Laboratory, University College London,
Holmbury St. Mary, Dorking, Surrey RH5 6NT, UK
}
\affil[e]{
Department of Physics, Kyoto University,
Kitashirakawa-oiwake-cho, Sakyo-ku, Kyoto 606-8502, Japan
}
\affil[f]{Astrobiology Center,
2-21-1 Osawa, Mitaka, Tokyo 181-8588, Japan
}
\affil[g]{
Tsung-Dao Lee Institute, Shanghai Jiao Tong University,
Shengrong Road 520, 201210 Shanghai, China
}
\affil[h]{
Astronomisches Rechen-Institut, Zentrum f\"{u}r Astronomie der Universit\"{a}t Heidelberg,
M\"{o}nchhofstr. 12-14, 69120 Heidelberg, Germany
} 
\affil[i]{The Institute of Statistical Mathematics,
10-3 Midori-cho, Tachikawa, Tokyo 190-8562, Japan
}
\affil[j]{Department of Marine Electronics and Mechanical Engineering, Tokyo University of Marine Science and Technology,
4-5-7 Konan, Minato-ku, Tokyo 108-8477, Japan
}
\affil[k]{
National Institute of Technology (KOSEN), Kochi College,
200-1 Monobe, Nankoku, Kochi 783-8508, Japan
}
\affil[l]{
Waseda Research Institute for Science and Engineering, Waseda University,
1-104 Totsukamachi, Shinjuku-ku, Tokyo 169-8050, Japan
}
\affil[m]{
Subaru Telescope Okayama Branch, National Astronomical Observatory of Japan,
3037-5 Honjo, Kamogata, Asakuchi, Okayama 719-0232, Japan
}
\affil[n]{
Astronomical Science Program, the Graduate University for Advanced Studies, SOKENDAI,
2-21-1 Osawa, Mitaka, Tokyo 181-8588, Japan
}

\authorinfo{Further author information: (Send correspondence to T.K.)\\T.K.: E-mail: kamizuka@ioa.s.u-tokyo.ac.jp}

\begin{document} 
\maketitle

\begin{abstract} 
{\it JASMINE} is a Japanese planned space mission that aims to reveal the formation history of our Galaxy and discover habitable exoEarths. For these objectives, the {\it JASMINE} satellite performs high-precision astrometric observations of the Galactic bulge and high-precision transit monitoring of M-dwarfs in the near-infrared (1.0–-1.6\,$\mu$m in wavelength). For feasibility studies, we develop an image simulation software named JASMINE-imagesim, which produces realistic observation images. This software takes into account various factors such as the optical point spread function (PSF), telescope jitter caused by the satellite's attitude control error (ACE), detector flat patterns, exposure timing differences between detector pixels, and various noise factors. As an example, we report a simulation for the feasibility study of astrometric observations using JASMINE-imagesim. The simulation confirms that the required position measurement accuracy of 4\,milliarcseconds for a single exposure of 12.5-mag objects is achievable if the telescope pointing jitter uniformly dilutes the PSF across all stars in the field of view. On the other hand, the simulation also demonstrates that the combination of realistic pointing jitter and exposure timing differences in the detector can significantly degrade accuracy and prevent achieving the requirement. This means that certain countermeasures against this issue must be developed. This result implies that this kind of simulation is important for mission planning and advanced developments to realize more realistic simulations help us to identify critical issues and also devise effective solutions.
\end{abstract}


\section{INTRODUCTION}
\label{sec:intro}

Distance to astronomical objects is a fundamental parameter for understanding their physical properties and 3-dimensional distribution, but it is difficult to measure. Parallax measurement, enabled by precise astrometric observations, is the most rigorous method of distance measurement. Two space missions have conducted such measurements across the whole sky. The first mission is {\it Hipparcos}, which measured parallaxes with an accuracy of ${\lesssim}1$\,milliarcseconds (mas) and allowed distance measurements of objects within a few hundred parsecs (pc)\cite{vanLeeuwen07a, vanLeeuwen07b}. The other mission is {\it Gaia}, which is still ongoing and has already achieved a parallax accuracy of up to ${\sim}20$\,microarcseconds ($\mu\mathrm{as}$), determining distances of ${\lesssim}10$\,kpc\cite{GDR3,GDR3E}. Their data enable precise quantitative studies of astronomical objects and detailed studies of the formation history of our Galaxy.

Even for the successful {\it Gaia} mission, measuring distances to objects with heavy interstellar and/or circumstellar extinctions is challenging, because {\it Gaia} observes objects in optical wavelengths where the extinctions make objects appear faint. Near-infrared (NIR) observation is a good way to overcome this problem and measure the distances to such objects. A NIR version of {\it Gaia} has already been proposed in Europe ({\it GaiaNIR})\cite{Hobbs19}. The Japanese community is also planning a NIR astrometric space mission named {\it Japan astrometry satellite mission for infrared exploration} ({\it JASMINE}; P.I.: N.~Gouda)\cite{Gouda20, Kawata24, Kataza24}. 

{\it JASMINE} is a planned M-class space mission by the Institute of Space and Astronautical Science (ISAS), Japan Aerospace Exploration Agency (JAXA). The satellite will be launched into a Sun-synchronous orbit with a sun shield pointing towards the Sun. In this orbit, the Galactic bulge is observable in Spring and Autumn, and the satellite will be dedicated to observing the Galactic bulge during these seasons. These observations enable astrometric observations of objects in the bulge where the heavy interstellar extinction has prohibited detailed studies by {\it Gaia} and to investigate the formation history of the nuclear star cluster, the nuclear stellar disk, and the bar structure that are key to understanding the formation history of the Galaxy. In the remaining seasons (i.e., Summer and Winter), the satellite will be dedicated to transit monitoring of M-type dwarfs to find habitable Earth-like exoplanets (exoEarths). Possible science cases other than these are also considered\cite{Kawata24}.

Observations are performed in a NIR band named $H_\mathrm{w}$ band, with wavelengths ranging from 1.0--1.6\,$\mu\mathrm{m}$, utilizing InGaAs detectors, and require high astrometric and photometric accuracies. As the full-success criteria, the astrometric observation needs to achieve a parallax accuracy of 40\,$\mu\mathrm{as}$ for objects brighter than $H_\mathrm{w}=12.5\,\mathrm{mag}$ and a proper motion accuracy of 125\,$\mu\mathrm{as/yr}$ for objects brighter than $H_\mathrm{w}=14.5\,\mathrm{mag}$, while 25\,$\mu\mathrm{as}$ and 25\,$\mu\mathrm{as/yr}$ are required as the parallax and proper motion accuracies for objects brighter than $H_\mathrm{w}=12.5\,\mathrm{mag}$ in the extra-success criteria, respectively. Assuming an accuracy improvement through ${\sim}67,000$ exposures for an object over the 3-year mission lifetime (improved by $\sqrt{67,000}$; see also Ref.~\citenum{Ohsawa24}) and some margins for unexpected error sources, a position measurement accuracy of 4\,mas is required for a single exposure of 12.5-mag objects. The transit photometry requires detecting 0.3 and 0.1\% dimming in objects with $H_\mathrm{w}=7.1\text{--}11.5$\,mag in the full- and extra-success criteria, respectively.

The feasibility of these requirements is critical for planning the mission, and we develop an observation simulation software named JASMINE-imagesim for feasibility evaluation. We report on the software and a simulation example for astrometric observations. Section~\ref{sec:imagesim} provides an overview of the software. Section~\ref{sec:simulation} describes the simulation settings, and Sec.~\ref{sec:results} presents the results. Finally, the summary is given in Sec.~\ref{sec:summary}.

\section{JASMINE-imagesim}
\label{sec:imagesim}

JASMINE-imagesim is a Python-based software\footnote{Available on \url{https://github.com/JASMINE-Mission/jasmine-imagesim}.} that simulates 2D images taken by the {\it JASMINE} satellite in the following way. The satellite is equipped with a telescope having an aperture diameter of 36\,cm and an effective focal length of ${\sim}4370\,\mathrm{mm}$ (see also Ref.~\citenum{Kataza24}). The focal plane has four InGaAs arrays (arranged in a 2$\times$2 grid) with a pixel pitch of 10\,$\mu\mathrm{m}$ and a resulting pixel scale of ${\sim}470\,\mathrm{mas/pix}$. The entire pixel format is 1968$\times$1968 pixels, while the photosensitive region is limited to the central 1952$\times$1952 pixels, corresponding to 15$\times$15\,$\mathrm{arcmin}^2$. Stars are imaged by the telescope onto the detector arrays, and the signals are read out by a 16-channel readout circuit in correlated double sampling (CDS) mode with rolling shutter readout (see also Sec.~\ref{sec:det}). Data for astrometric observations are obtained in a step-stare sequence (see also Ref.~\citenum{Ohsawa24}): 1) multiple images are obtained in staring mode (i.e., the satellite is controlled such that the telescope stares at a fixed point); 2) the telescope is moved (stepped) by half the field of view (FoV); 3) multiple images are obtained again; 4) the telescope is moved to another point in the survey region. The exposure time is set to 12.5\,seconds. In the transit observations, targets are constantly imaged in staring mode with longer exposure times depending on the brightness of each target.

Several factors affect the obtained images in the above process, such as the optical processes that create the point spread function (PSF), telescope pointing jitter caused by the attitude control error (ACE) of the satellite, integration and readout method of the detector, etc. JASMINE-imagesim is developed to simulate realistic observation images by taking these effects into account.

\subsection{Optical PSF Calculation}
\label{sec:psf}

\begin{figure}[bt]
\begin{center}
\begin{tabular}{c} 
\includegraphics[height=6cm]{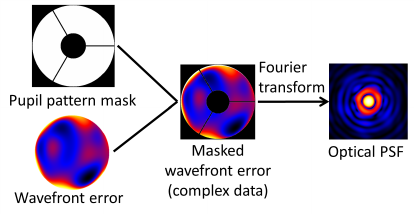}
\end{tabular}
\end{center}
\caption{
\label{fig:psf}
Conceptual diagram of the optical PSF calculation.}
\end{figure} 

In the simple case without wavefront errors, the optical PSF created by the telescope can be calculated through the Fourier transform of the pupil pattern shown in Fig.~\ref{fig:psf}. The pupil pattern of the {\it JASMINE} telescope is a circular aperture with obstructions caused by the secondary mirror and its supporting spiders arranged in a Y-shaped pattern. The outer diameter is defined by the aperture diameter of the telescope (36\,cm), the diameter of the central obstruction is 12.6\,cm, and the thickness of the spiders is 5\,mm.

To simulate realistic optical PSFs, the wavefront error should be considered in the simulation. Wavefront errors are often expressed by Zernike polynomials, and JASMINE-imagesim accepts data containing the 37 Zernike coefficients in the Fringe convention calculated by optical design software like CODE~V and Zemax (Ref.~\citenum{Niu22} and references therein). Position dependence can also be handled in the simulation. The wavefront error pattern is created by this function and masked by the pupil pattern (Fig.~\ref{fig:psf}). The optical PSF is calculated by performing the Fourier transform on the resulting pattern. Since the wavelength dependence of the PSF is also an important factor in simulating images, this calculation is done at each wavelength grid defined in the settings for the simulation.

\subsection{ACE Calculation}
\label{sec:ace}

\begin{figure}[tb]
\begin{center}
\begin{tabular}{c} 
\includegraphics[height=5.5cm]{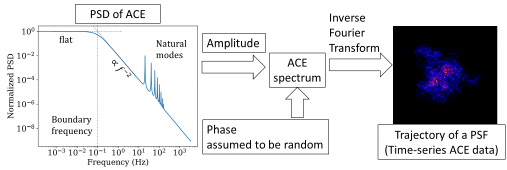}
\end{tabular}
\end{center}
\caption{
\label{fig:ace}
Conceptual diagram of the ACE calculation.}
\end{figure} 

Satellites have ACE to some extent, which causes pointing jitter of the telescope and wobbles the stellar image on the focal-plane detector. Since such a wobble blurs the resulting images, the effect of the ACE should be incorporated into the simulation.

The characteristics of ACE vary from satellite to satellite and can be expressed by the power spectral density (PSD; Fig.~\ref{fig:ace}). The shape of PSDs usually consists of power-law components and some peak components, which are characterized by the attitude control system and the satellite structure (i.e., natural modes), respectively. In JASMINE-imagesim, the PSD is assumed to consist of a flat (white) component in the low-frequency regime, a power-law component with a power-law index of $-2$ in the high-frequency regime, and natural mode components expressed by Lorentzian functions. Based on parameters defining the PSD shape, the simulator creates the Fourier components of the ACE by adding some noise in the amplitude and setting random phase values and then calculates a time-series dataset of the ACE through the inverse Fourier transform. The resulting ACE data are finally scaled such that their standard deviation equals the value specified in the input parameters.

\subsection{Integration and Readout}
\label{sec:det}

Now that we have the data of the optical PSF and the wobbling motion of the image caused by the ACE from the above processes, we can simulate the temporal variation of the signal distribution on the detector for each star. If the starting and finishing times of the exposure for each pixel are defined, we can calculate the integrated signal by integrating the varying signal during the exposure.

\begin{figure}[tb]
\begin{center}
\begin{tabular}{c} 
\includegraphics[height=6.5cm]{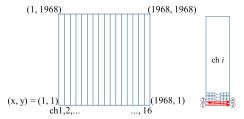}
\end{tabular}
\end{center}
\caption{
\label{fig:det}
(Left) Detector structure and (Right) the readout sequence. The array consists of 16 vertical stripe areas. Each area is read by a different readout circuit. Within the area, the reset and readout sequence starts from the left bottom corner (mark 1) and goes in the row direction. After reaching the right edge (mark 2), that sequence goes to the left edge of the next row (mark 3) and goes in the row direction again (mark 4).}
\end{figure}

In the {\it JASMINE} mission, the detector array is read out by 16 readout channels, and the area read out by each readout channel is a vertical rectangular area of 123$\times$1968~pixels aligned in the row direction (Fig.~\ref{fig:det}). These areas are read and reset in parallel, and in each area, the read and reset of the pixels are sequentially performed from bottom to top in the order shown in Fig.~\ref{fig:det}. Therefore, the exposure duration is the same for all pixels, but the timing of the start and end of exposure differ from pixel to pixel depending on the position on the detector. JASMINE-imagesim is designed to allow this effect to be incorporated.

In addition, sensitivity inhomogeneity within a pixel (i.e., intrapixel flat) can also be considered in the simulation. This function is achieved by performing the integration process by dividing a pixel into subpixels and multiplying the resulting signal pattern by the intrapixel flat pattern.

The integration calculation is accelerated by using a Graphics Processing Unit (GPU) because the integration calculation for each subpixel can be performed in parallel. This process is performed for each star due to the limited number of threads. After the integration calculation, signals for each pixel are obtained by summing up the signals of the subpixels, and the resulting pixelated images are embedded in the entire image of the detector.

As a final step, the pattern of the sensitivity differences between pixels (i.e., interpixel flat pattern) is multiplied to the resulting image, additional signals such as dark current, background signal, shot noise, and readout noise are added, and then digitization is performed to obtain the final image.

\section{Simulation and Analysis}
\label{sec:simulation}
The feasibility of achieving the requirements can be investigated through simulations using JASMINE-imagesim. Here, we present an example of a simulation for astrometric observations and the evaluation of the position measurement accuracy of stellar images.

For the simulation, we produce 11 images with 400 stars distributed in a grid-like pattern (a 20$\times$20 grid). The interval between the stars is 98.4\,pixels to space them equally. However, since the stellar positions relative to the pixel grid should be random, random fluctuations are added to the intervals. The brightness of the stars is set to be identical to investigate the brightness dependence of the position measurement accuracy. The color of the stars is also set to be identical. The major parameters are shown in Table\,\ref{tab:params} 

To evaluate the effect of the ACE and the exposure time difference across the detector, we create two types of images with different assumptions about the ACE. The first case is that the ACE profile integrated during the exposure can be expressed by a Gaussian profile and is identical for all stars (Gaussian ACE case). The other case is that the ACE is calculated to simulate a realistic case and the integrated profile can vary between pixels due to the exposure timing differences (realistic ACE case). For the latter case, the PSD shown in Fig.~\ref{fig:ace} is used for that of the ACE (i.e., boundary frequency is 0.1\,Hz; eight normal modes in 20--160\,Hz). The standard deviation of the ACE in the Gaussian ACE case is set to be 300\,mas, expected for {\it JASMINE}, and in the realistic ACE case, the calculated ACE is scaled such that the median standard deviation of the ACE in the 12.5-sec exposure equals 300\,mas.

The resulting images are analyzed using the effective PSF (ePSF) method\cite{Anderson00} to evaluate the position measurement accuracy. The ePSF is a PSF defined in pixel coordinates taking the effects of pixelation and the intrapixel flat into account, and it differs from the intrinsic optical PSF. This method derives the ePSF from the image and estimates the stellar positions by fitting the obtained ePSF to the images. These processes are iterated in the analysis, and the number of iterations is set to three in this study. The position measurement accuracy is evaluated as the standard deviation of the difference between the estimated stellar positions and the true positions. The consideration of ePSF is especially important in cases where the PSF is undersampled. For {\it JASMINE}, the intrinsic PSF has a full-width at half-maximum (FWHM) of 740\,mas, corresponding to 1.6\,pixels, which is not very sufficient for accurate sampling. Therefore, it is preferable to use this type of analysis.

\begin{table}[tb]
\caption{Parameters assumed in the simulation.}
\label{tab:params}
\begin{center}       
\begin{tabular}{ll} 
\hline\hline
\rule[-1ex]{0pt}{3.5ex}  Name & Value  \\
\hline
\rule[-1ex]{0pt}{3.5ex}  Telescope aperture diameter & 36\,cm   \\
\rule[-1ex]{0pt}{3.5ex}  Diameter of the secondary mirror & 12.6\,cm \\
\rule[-1ex]{0pt}{3.5ex}  Spider thickness & 5\,mm\\
\rule[-1ex]{0pt}{3.5ex}  Effective focal length & 4369\,mm\\
\rule[-1ex]{0pt}{3.5ex}  Dark current and stray light & 24.5\,e$^-$/pix/sec\\
\rule[-1ex]{0pt}{3.5ex}  Readout noise & 15\,e$^-$/read\\
\rule[-1ex]{0pt}{3.5ex}  Conversion gain & 3\,e$^-$/adu \\
\rule[-1ex]{0pt}{3.5ex}  Readout speed & 5\,$\mu\mathrm{s/pix}$ (200-kHz sampling)\\
\rule[-1ex]{0pt}{3.5ex}  Interpixel flat & 1\% (1$\sigma$)\\
\rule[-1ex]{0pt}{3.5ex}  Exposure time & 12.5\,seconds/image\\
\rule[-1ex]{0pt}{3.5ex}  $H_\mathrm{w}$ magnitude & 12.0, 12.5, $\dotsc$, 14.5\,mag\\
\rule[-1ex]{0pt}{3.5ex}  $J{-}H$ color index & 2.0\,mag\\
\hline \hline
\end{tabular}
\end{center}
\end{table}

\section{Results}
\label{sec:results}

\begin{figure}[b]
\begin{center}
\begin{tabular}{c} 
\includegraphics[height=7cm]{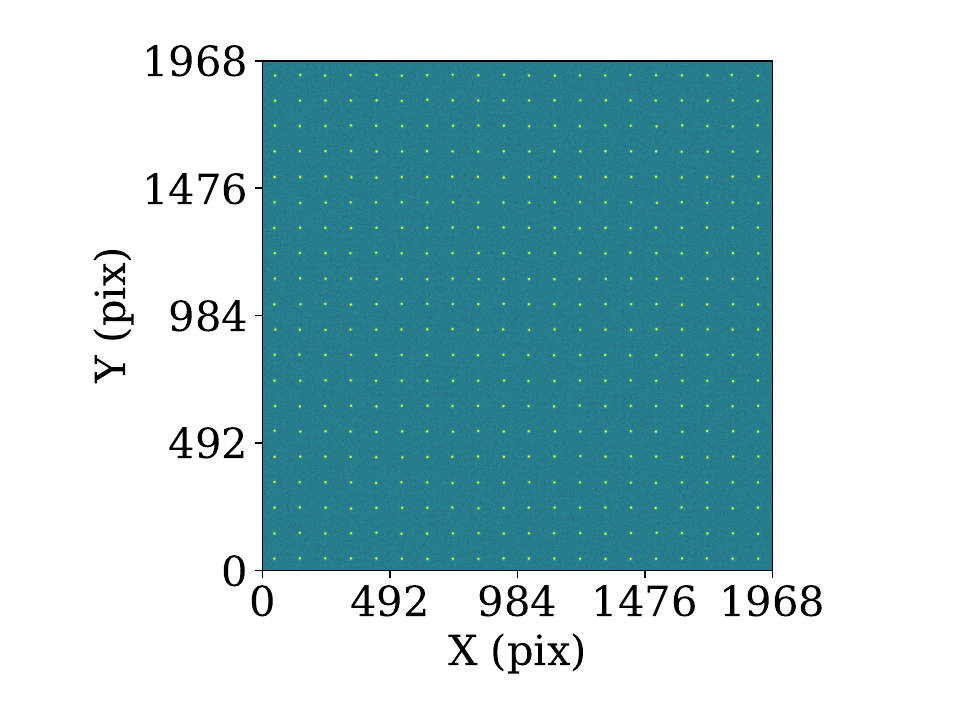}
\end{tabular}
\end{center}
\vspace{-0.5cm}
\caption{
\label{fig:full}
Simulated full image for 12.0-mag objects.}
\end{figure} 

\begin{figure}[bt]
\begin{center}
\begin{tabular}{c} 
\hspace{1cm}
\includegraphics[height=7cm]{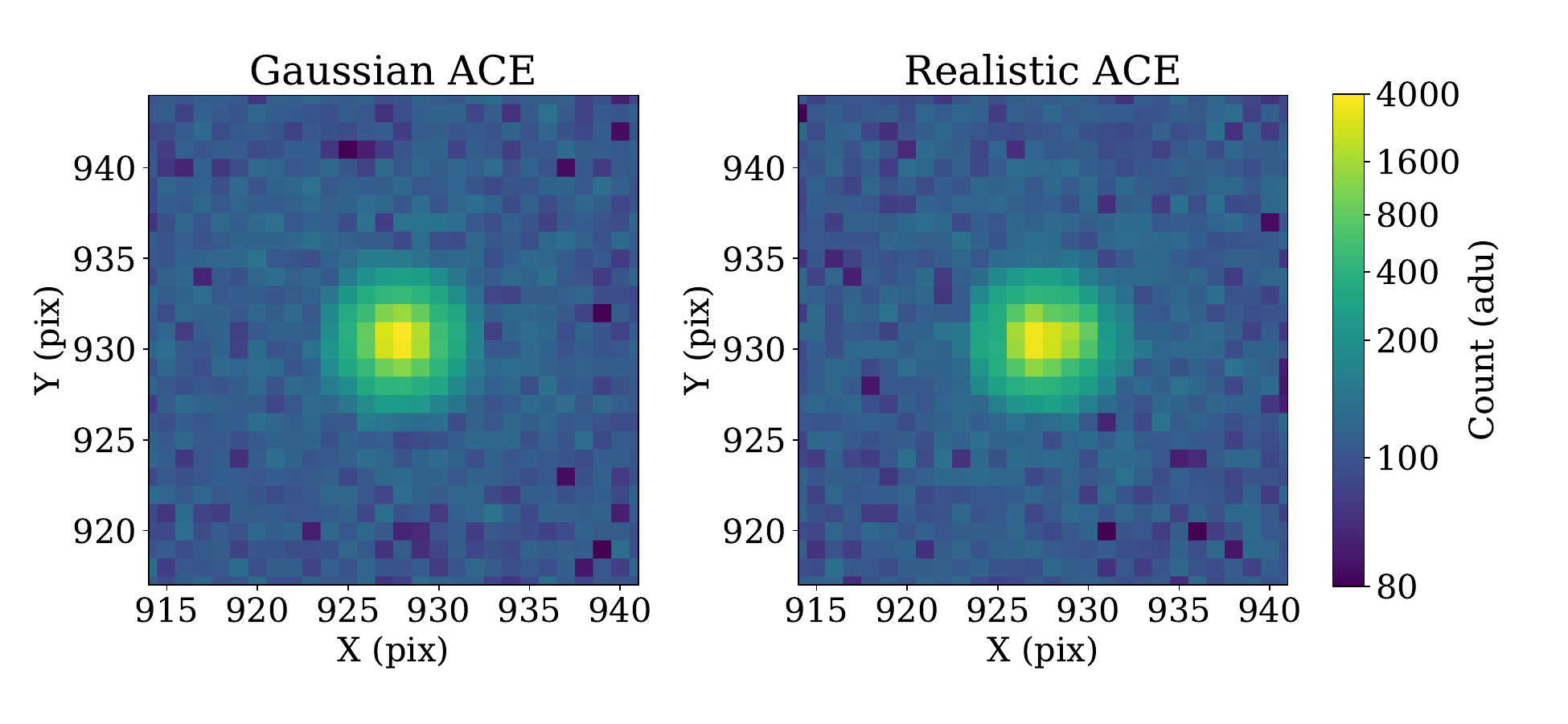}
\end{tabular}
\end{center}
\vspace{-0.5cm}
\caption{
\label{fig:zoom}
Close-up image of a 12.0-mag star in the Gaussian ACE case (left) and the realistic ACE case (right).}
\end{figure} 

Figure\,\ref{fig:full} shows a full image simulating 12.0-mag stars. We can confirm that 400 stars are distributed in a grid pattern as intended. Figure\,\ref{fig:zoom} shows close-up images of a star among the 400 stars. The left and right panels are for the Gaussian ACE case and the realistic ACE case, respectively. In the Gaussian ACE case, the stellar image exhibits an almost point-symmetric image, while a slightly extended asymmetric image is seen in the realistic ACE case, although both are the same star imaged in the same conditions except for the ACE assumption. This indicates that the ACE can distort and dilute the stellar image. 

\begin{figure}[bt]
\begin{center}
\begin{tabular}{c} 
\hspace{-1cm}
\includegraphics[height=7.5cm]{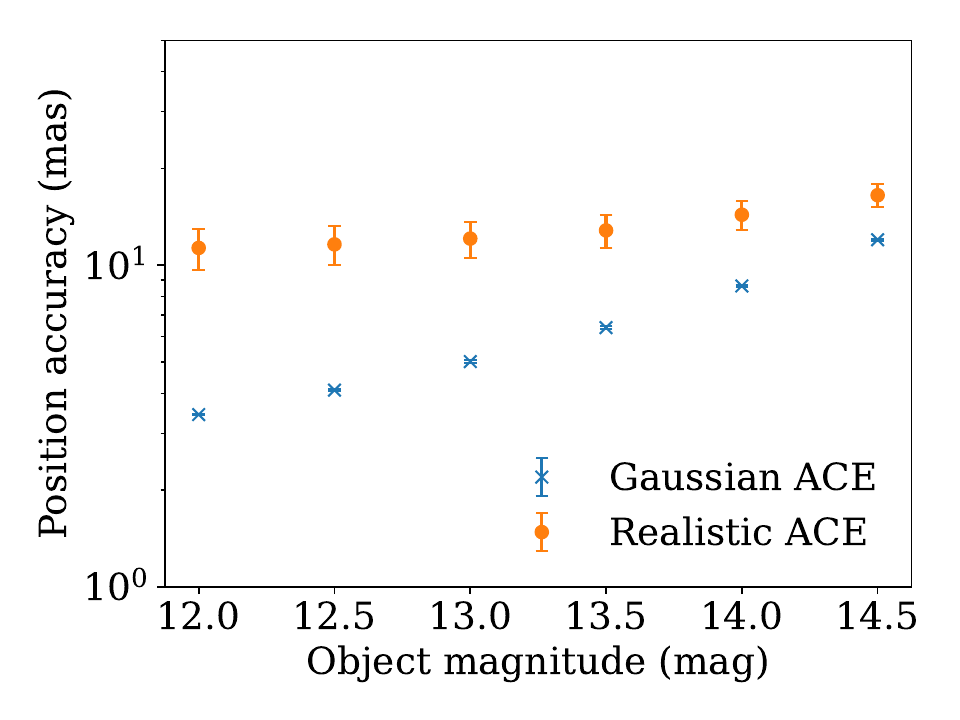}
\end{tabular}
\end{center}
\vspace{-0.5cm}
\caption{
\label{fig:acc_comp}
Position measurement accuracy for a single exposure data against object magnitude.}
\end{figure} 

\begin{figure}[bt]
\begin{center}
\begin{tabular}{c} 
\hspace{1cm}
\includegraphics[height=8cm]{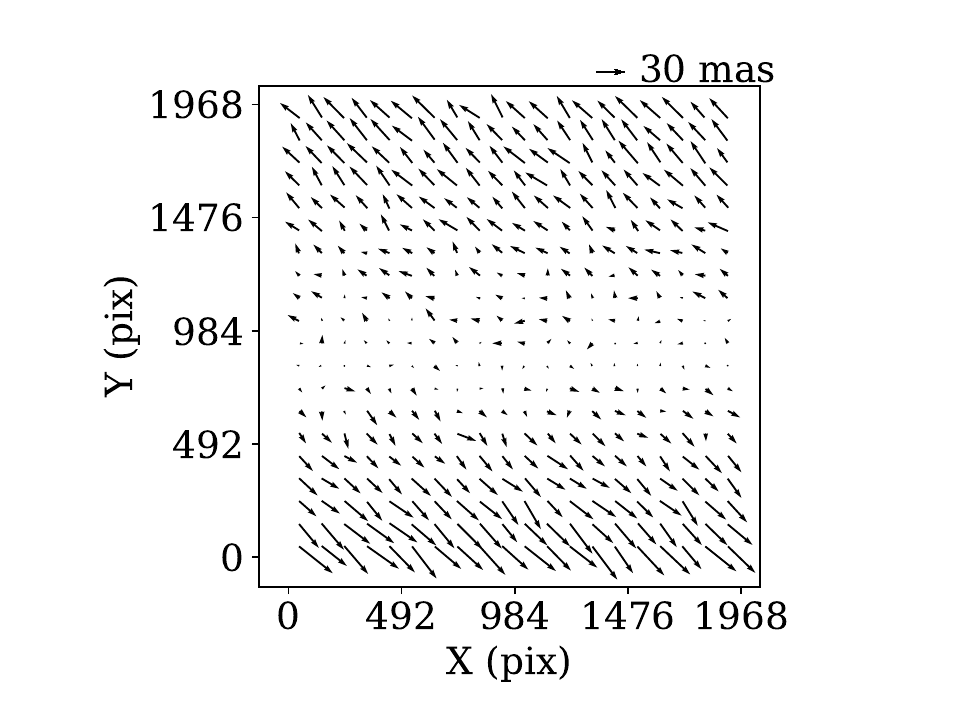}
\end{tabular}
\end{center}
\vspace{-0.5cm}
\caption{
\label{fig:deviation}
Vector plot of the deviation of the estimated position from the true position obtained in the 12.0-mag simulation in the realistic ACE case. The scale reference is shown in the top right corner.}
\end{figure} 

Figure\,\ref{fig:acc_comp} shows the evaluated position measurement accuracy for a single exposure data. The accuracies are evaluated for each image, and 11 values are obtained from one dataset (11 images). The plotted values are the mean of these 11 values, and the error bar shows the standard error of the mean. The overall tendency is that the accuracy decreases as the object becomes fainter. This tendency can be simply understood as the signal-to-noise ratio decreases in the faint regime. We can also confirm that the accuracy for 12.5-mag objects in the Gaussian ACE case is about 4\,mas, and the accuracy requirement can be achieved in this case. However, in the realistic ACE case, the accuracy for 12.5-mag objects is degraded to ${\sim}10\,\mathrm{mas}$. This means that the achievement of the required accuracy can be difficult under the realistic ACE. 

Figure\,\ref{fig:deviation} shows a vector plot of the deviation of the estimated position from the true position. This is for 12.0-mag objects in the realistic ACE case and shows a clear systematic pattern. Such a pattern does not appear in the Gaussian ACE case. In addition, this pattern has a trend in the vertical direction. Based on these characteristics, the systematic deviation pattern is thought to be caused by the combination of the exposure timing difference appearing in the vertical direction and the telescope jitter. If the telescope jitter exhibits a low-frequency movement, and such a jitter occurs around the start and/or end timing of exposure, PSF shapes will differ in the vertical direction, and analysis assuming a uniform ePSF will lead to a systematic error as seen in this experiment due to the systematic difference in ePSF shape. Such events occur stochastically, and the fraction of accurate observation can significantly decrease. This simulation shows that countermeasures against this problem are important for the achievement of the mission.

\section{Summary}
\label{sec:summary}
{\it JASMINE} is a Japanese planned space mission aiming to conduct astrometric observations and transit monitoring in the NIR. Its objectives are to study the history of the inner Galactic structures and to search for exoEarths. To assess the feasibility of achieving the required high astrometric and photometric accuracies, we developed an observational image simulator named JASMINE-imagesim. This simulator generates realistic images by considering various factors that influence image quality, such as realistic PSF, telescope jitter, exposure timing of each detector pixel, intra- and inter-pixel flat patterns, etc.

We reported an example of astrometric observation simulation and the evaluation of position measurement accuracy. The simulation produced images in the Gaussian ACE case and the realistic ACE case. Position measurement accuracy was evaluated through the ePSF analysis, and it was found that the combination of the ACE and the exposure timing differences in the detector can significantly degrade the accuracy, thus hindering the achievement of the requirement. This means that we have to develop countermeasures to address this issue.

As potential countermeasures, the implementation of global reset mode and sample hold mode to the detector is promising, because they can eliminate the exposure time differences. Reducing the ACE is another option. However, these hardware measures are costly. We are also investigating analysis methods to deal with the variety of ePSFs across the detector, such as the following process: dividing the analysis region into four regions along the Y-axis, constructing representative ePSFs for each region, and calculating ePSFs for each Y-axis position by interpolation. Since the systematic deviation pattern shown in Fig.~\ref{fig:deviation} depends on the Y-axis position, this approach may help mitigate the issue caused by the ACE. JASMINE-imagesim will be used to evaluate these methods and help to find an effective solution.

Additional factors, such as the crowded stellar distribution towards the Galactic bulge, large-scale patterns caused by diffuse emission and interstellar extinction, and the coexistence of stars with various brightnesses, may also impede mission success. We recently updated JASMINE-imagesim to incorporate these factors and generate mock Galactic center images. We are investigating the performance of the mission with more realistic simulations.


\acknowledgments 
This work made use of Astropy:\footnote{http://www.astropy.org} a community-developed core Python package and an ecosystem of tools and resources for astronomy \cite{astropy:2013, astropy:2018, astropy:2022}.

\bibliography{report}
\bibliographystyle{spiebib} 

\end{document}